\documentclass[twocolumn, prb,showpacs,preprintnumbers,amsmath,amssymb,superscriptaddress]{revtex4}% Physical Review B

\usepackage{graphicx}% Include figure files
\usepackage{dcolumn}% Align table columns on decimal point
\usepackage{bm}% bold math
\usepackage{textcomp} 
\usepackage{color}
\usepackage{natbib}
\usepackage{amsmath, amsfonts, amssymb}
\usepackage{enumerate}
\usepackage[breaklinks]{hyperref}
\usepackage[hyphenbreaks]{breakurl}
\usepackage{paralist}

\begin{document}

%\preprint{PRB}

\title{Transport spectroscopy of induced superconductivity in the three-dimensional topological insulator HgTe}% Force line breaks with \\

\author{Jonas Wiedenmann}
\email{Jonas.Wiedenmann@physik.uni-wuerzburg.de}
\affiliation{%
Experimentelle Physik III, Physikalisches Institut, Universit\"at W\"urzburg, Am Hubland, D-97074 W\"urzburg, Germany
}%

\author{Eva Liebhaber}%
\affiliation{%
Experimentelle Physik III, Physikalisches Institut, Universit\"at W\"urzburg, Am Hubland, D-97074 W\"urzburg, Germany
}%
\author{Johannes K\"ubert}
\affiliation{%
Experimentelle Physik III, Physikalisches Institut, Universit\"at W\"urzburg, Am Hubland, D-97074 W\"urzburg, Germany
}%
\author{Erwann Bocquillon}
\affiliation{%
Experimentelle Physik III, Physikalisches Institut, Universit\"at W\"urzburg, Am Hubland, D-97074 W\"urzburg, Germany
}%
\affiliation{Laboratoire Pierre Aigrain, D\'{e}partement de physique de l\textquotesingle ENS, \'{E}cole Normale Sup\'{e}rieure, PSL Research University, Universit\'{e} Paris Diderot, Sorbonne Paris Cit\'{e}, Sorbonne Universit\'{e}s, UPMC Univ. Paris 06, CNRS, 75005 Paris, France}
\author{Pablo Burset}
\affiliation{Department of Applied Physics, Aalto University, FIN-00076 Aalto, Finland}
\author{Christopher Ames}
\affiliation{%
Experimentelle Physik III, Physikalisches Institut, Universit\"at W\"urzburg, Am Hubland, D-97074 W\"urzburg, Germany
}%
\author{Hartmut Buhmann}
\affiliation{%
Experimentelle Physik III, Physikalisches Institut, Universit\"at W\"urzburg, Am Hubland, D-97074 W\"urzburg, Germany
}%
\author{Teun M. Klapwijk}

\affiliation{
Kavli Institute of Nanoscience, Faculty of Applied Sciences,
Delft University of Technology, Lorentzweg 1, 2628 CJ Delft, The Netherlands
}%
\author{Laurens W. Molenkamp}
\affiliation{%
Experimentelle Physik III, Physikalisches Institut, Universit\"at W\"urzburg, Am Hubland, D-97074 W\"urzburg, Germany
}%

\date{\today}% It is always \today, today,

\begin{abstract}
The proximity-induced superconducting state in the 3-dimensional topological insulator HgTe has been studied using electronic transport of a normal metal-superconducting point contact as a spectroscopic tool (Andreev point contact spectroscopy). By analyzing the conductance as a function of voltage for various temperatures, magnetic fields and gate-voltages, we find evidence, in equilibrium, for an induced order parameter in HgTe of $70\,$\textmu eV and a niobium order parameter of $1.1\,$meV. To understand the full conductance curve as a function of applied voltage we suggest a non-equilibrium driven transformation of the quantum transport process where the relevant scattering region and  equilibrium reservoirs change with voltage. This implies that the spectroscopy probes the superconducting correlations at different positions in the sample, depending on the bias voltage. 
%Inducing superconducting pairing into the surface states of a topological insulator is predicted to lead to the emergence of mixed spin singlet/triplet superconducting correlations and Majorana zero modes.Here, the three-dimensional topological insulator strained HgTe is proximitized with the conventional s-wave superconductor niobium and studied via point contact Andreev reflection spectroscopy. The differential conductance of the devices is measured and shows a strong zero bias peak and clear signatures of the superconducting gap. The Blonder-Tinkham-Klapwijk formalism provides an appropriate framework to analyse the measurements, and we identify two different energy scales, associated to two barrier strengths. One is rather large ($Z_2 \approx 1.1$) at the superconductor-topological insulator interface while a smaller one ($Z_1 \approx 0.4$) is located  at the point contact. We explain our measurements in the context of non-equilibrium superconducting transport and furthermore investigate the influence of temperature, electron density and magnetic field.
\end{abstract}

\maketitle

\section{\label{sec:level}Introduction\protect}

The two most important methods to obtain reliable quantitative spectral information about the electronic properties of a superconductor are Giaever tunneling \cite{Wolf2011} and point contact Andreev spectroscopy \cite{Blonder1982b, Daghero2010}. In tunnel spectroscopy two metal thin films are weakly coupled by an insulating tunnel barrier, leading to a current-voltage characteristic which is controlled by the unperturbed superconducting densities of states in both metals, $N_s(E)$, and their occupation, given by the Fermi-functions, $f_0(E)$. The technique can also be used successfully to study the proximity-effect in superconducting bilayers as experimentally shown by Wolf and Arnold \cite{WOLF1982}, but requires the difficult development of an opaque tunnel barrier. The second method, point contact Andreev spectroscopy, has become a standard tool to evaluate the microscopic properties of new bulk materials. The experimental configuration consists of a macroscopically sized point-shaped metal wire, which touches a superconducting material, usually a single crystal. In the contact area the conductance in both the superconducting and normal regime is dominated by the channels with the highest transmission usually loosely called 'pinholes'. Thus, there is no need to know the exact nature of the contacting layer and the transmissivity of the point contact %modelled in the BTK model by the barrier parameter $Z$
can be assumed to reach values in the order of one, without disturbing the properties of the superconductor. This latter assumption is valid because the two bulk materials are connected by an area which is very small compared to the lateral dimensions of the materials and assumed to be smaller than the elastic mean free path both materials (ballistic transport). Such a geometry leaves the reservoirs undisturbed, a crucial condition for the determination of the electronic parameters of the superconductor and generalized in the Landauer-B\"uttiker picture of quantum transport.   

{Our aim in this paper is to apply Andreev spectroscopy to the proximity-induced superconducting state in a 3D topological insulator (3DTI). The application of Andreev-spectroscopy to low dimensional heterostructures is a much less mature experimental technique than for bulk systems. The point contact has to be lithographically defined and is therefore usually larger than for bulk systems, where accidentally formed pinholes of smaller dimensions dominate the transport.   % In addition, the reservoirs constitute a much smaller number of electrons and are usually 2-dimensional. 
These experimental concerns are exacerbated in the case of spectroscopy on proximity-induced superconductivity, because of the need to use two dissimilar materials and, unavoidably, a complex lithographically structured geometry. %One constraint is the transmissivity between the main superconductor (S$_\textrm{m}$) and the material in which the superconducting state is induced. In addition, the geometry to which the induced superconductivity is confined needs to be known and controlled. 
In fact, very few successful spectroscopic experiments on proximitized systems have been carried out. One example, on diffusive systems, is by Scheer $et~al.$  \cite{Scheer2000}, using mechanical break junctions, an approach that merges bulk point contact behavior with thin films. Recently, Kjaergaard $et~al.$ \cite{Kjaergaard2016} have presented results on point contact spectroscopy in the ballistic Al/InAs system, which partially fulfills the experimental requirements. It shows the expected doubling of the quantized conductance steps for point contacts in the highly transmissive regime, but exhibits also, from a spectroscopic perspective, many puzzling results  and, additionally, unexpected behavior as a function of the tunable point contact transmissivity. A different geometry was used by Zhang $et~al.$ \cite{Zhang2016b}, also employing a tunable point contact, predominantly in the regime of low transmission.} 
       
\begin{figure}
\includegraphics[scale=0.6]{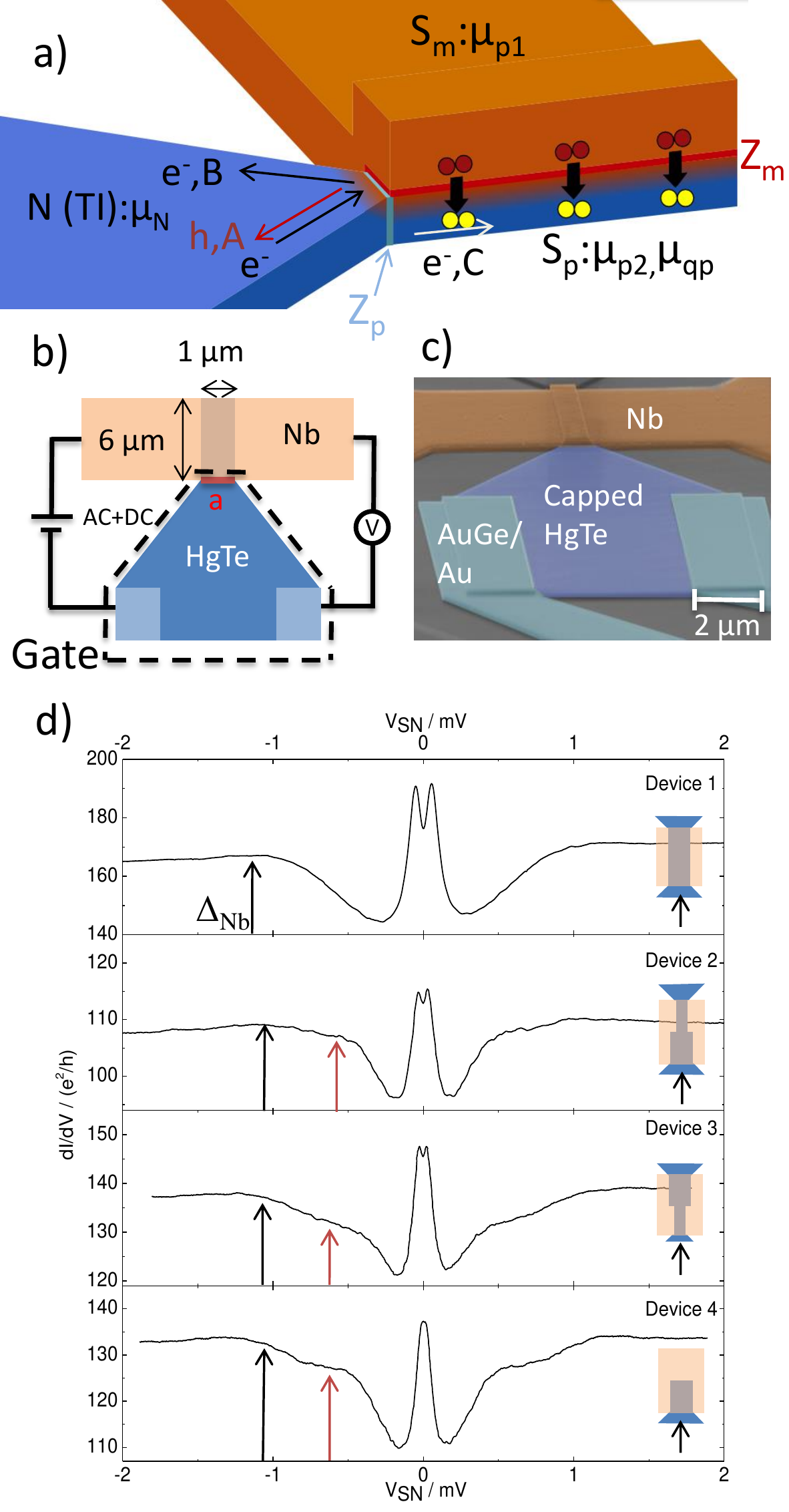}
\caption{a) Schematic of the experiment: A s-wave superconductor S$_{\textrm{m}}$ (orange) is inducing superconducting pairing in an underlying topological insulator S$_{\textrm{p}}$. This state is probed via a point contact. An electron impinging from the 3DTI reservoir N can either be Andreev reflected, normal reflected or transmitted with probability amplitudes A,B,C respectively. The current is carried away at the right of $Z_{\textrm{p}}$ as a supercurrent. b) Schematic of the device and measurement setup. A niobium strip is covering a HgTe bar which is coupled to an equilibrium reservoir via a small orifice marked with the letter 'a'. The dashed lines mark the contours of the gate. c) False color SEM picture of a device without a gate electrode. d) $dI/dV$ measurements of four devices. The devices differ by the 'connectivity' of the HgTe bar, covered by niobium, as indicated in the inset.}
\label{Fig:fig1}
\end{figure}

We report on a study of a high quality 3-dimensional topological insulator, epitaxially grown strained HgTe, which is proximitized by a conventional superconductor, niobium. In previous experiments we reported on the observation  of a 'missing $n=1$' Shapiro step \cite{Wiedenmann2016a}, an indication of an unconventional Josephson effect in 3DTI HgTe based Josephson junctions. The same type of observation was subsequently done in Josephson junctions in a 2D topological insulator showing a sequence of even-only Shapiro steps (up to n=10) and emission at half the Josephson frequency. Both signatures indicate at least a fractional $4\pi$-periodic Josephson effect, and point towards the presence of gapless Majorana-Andreev bound states \cite{Bocquillon2016b, Deacon2016}. %To account for these observations (see also\cite{Dominguez2017} it has been rewarding to assume that both $\sin\phi$ as well as $\sin{\phi/2}$ Josephson currents are present.
Since the Josephson effect arises from the proximity-induced superconducting state, we are interested in a determination of the energy dependent properties of this induced superconducting state, which in principle serves as a coherent reservoir for the Josephson effect, analogous to the established proximity-effect based niobium superconductor-insulator-superconductor (SIS) junctions \cite{Gurvitch1983}. It is crucial to be able to measure these electronic states directly, in particular because the Josephson-effect itself contains only information about the phase difference and the nature of the current-phase relation, but not about its energy dependence. For this reason we designed an experiment which is based on a NcS$_\textrm{p}$ point contact to emulate Andreev-spectroscopy of the induced superconducting state (N is a normal reservoir, which in our case is a topological insulator, c is the constriction, and S$_\textrm{p}$ is the proximity-induced superconductor), as schematically shown in Fig. \ref{Fig:fig1}a). % This geometry avoids the complicated dynamics of energy levels in a voltage-biased ScS configuration. 
Therefore, the strained HgTe is defined lithographically to a finite sized bar and covered over a small distance by a conventional superconductor S$_\textrm{m}$. We assume that an induced superconducting state exists underneath the superconducting material, which we label S$_\textrm{p}$. The electronic states in this region are the source for the observed Josephson effect. Note that in such a geometry no Majorana zero modes are expected to emerge due to the lack of confinement \cite{Snelder15b} but unconventional superconducting correlations might be observable \cite{Burset2014, Burset2015}.
We find that the electronic transport between the N-reservoir and the S$_\textrm{m}$ reservoir is governed by two energy scales which we identify as the superconducting gap of the niobium film $\Delta_{\textrm{Nb}}$ and the induced gap in the surface states of the HgTe, labeled $\Delta_{\textrm{p}}$. By using modeling as introduced by Blonder $et~al.$ \cite{Blonder1982b} we are able to show that the transmissivity at the HgTe/Nb interface is rather low. We argue that the voltage-carrying state, needed to obtain spectroscopic information, leads to a non-equilibrium occupation of the proximity-induced superconducting state, rendering the device into different experimental conditions, depending on the bias voltage. % in which the niobium superconducting gap is shown.  

%The paper is structured as follows. First, the design and fabrication of the point contact are discussed, then measurements of the conductance at zero magnetic field and zero gate voltage are shown. Subsequently,  measurements for variable gate and magnetic field are discussed. 

\section{\label{sec:level2}Sample description}

The NcS junctions in this work are based on epitaxially grown layers of strained HgTe sandwiched between Hg$_{0.3}$Cd$_{0.7}$Te capping layers. These additional layers have a conventional band structure and protect against surface oxidation, which reduces the carrier mobility. They also protect the strained HgTe during subsequent lithographic processing. The HgTe sandwich is shaped as a $1\,$\textmu m wide bar which at one or both ends tapers out with an angle of about $45^\circ$. The top Hg$_{0.3}$Cd$_{0.7}$Te capping layer is removed by dry etching and subsequently covered by niobium, which is in contact with the strained HgTe. Fig. \ref{Fig:fig1}a)  shows a schematic drawing of the device. % and Fig.\ref{Fig:fig1}c a false colour SEM image of a finished structure.
The orange part is the source-superconductor, S$_\textrm{m}$, made of niobium and the blue part is the strained HgTe. At the interface we allow for a finite transmission coefficient which is labeled $Z_{\textrm{m}}$. This dimensionless barrier is in general connected to the normal state transmission by $t=(1+Z^2)^{-1}$. %The processing used to fabricate this contact may also reduce the elastic mean free path in this part of the HgTe compared to the starting material. 
The superconducting correlations are induced in the HgTe indicated by yellow dots. The tapered part of the HgTe, not covered by the niobium, is left capped by the Hg$_{0.3}$Cd$_{0.7}$Te layer, and we assume that this part has the same mobility as the starting material. At the constriction we allow for an additional elastic scattering parameter $Z_{\textrm{p}}$.  

The quality of the HgTe layers is characterized using a Hall bar fabricated from the same wafer. At zero gate voltage ($V_g=0$) a density of $n_{\textrm{2D}} \approx 5 \times 10^{11} \, \textrm{cm}^{-2}$ and mobilities of $\mu \approx 200 \, 000 \, \textrm{cm}^2 /\textrm{Vs}$ are routinely achieved resulting in a mean free path $l_{\textrm{mfp}}\approx 2-3 \,$\textmu m. The mobility is about ten times lower when tuning the device into the p-regime. As shown in  detailed magnetotransport studies \cite{Brune2011,Brune2014}, clear quantum Hall plateaus are observed indicating transport mediated predominately by two dimensional states which were shown to originate from the topological surface states. 

The point  contact is fabricated using electron beam lithography and PMMA resist. As HgTe is sensitive to temperatures above $90 ^{\circ}$C, all bake-out and lift-off procedures are carried out well below this temperature. In a first step the HgTe mesa is defined using low energy argon sputtering. During this process a thin titanium etch shield, separated by a SiO$_2$ sacrificial layer from the HgTe, is protecting the mesa. The shield is afterwards removed by a buffered oxide etch dip.  The dimensions of the mesa as shown in Fig. \ref{Fig:fig1}b) and c) are chosen such that the orifice ($a=1 \, $\textmu m for Device 1, 2 and 4 and $a=0.6 \,$\textmu m for Device 3, respectively)  is smaller than the ballistic mean free path of the surface states. The size of the normal reservoir is much larger than this length scale, to allow full energy relaxation in this region. % The HgTe bar then opens widely with a $45 ^\circ$ angle. % to avoid scattering and to form a equilibrium reservoir for electrons. 
In a next step, the superconductor is deposited. %This step is crucial as it determines the quality of the interface between the 3DTI and the SC. 
Since the interface is buried the cap layer needs to be removed, which is done by argon etching, followed by in-situ magnetron sputtering of about $110 \,$nm of niobium. % A large superconducting equilibrium reservoir right after the constriction is used to avoid heating effects in the point contact. 
After this the leads for the Ohmic contacts are defined and $50\,$nm AuGe/$50\,$nm Au is deposited. The contact resistances are usually small ($<50 \, \Omega$).
To allow control of the charge carrier density in the 3DTI a top gate electrode is evaporated on top of the HgTe (c.f. dashed lines in Fig. \ref{Fig:fig1}b), as follows. First, a thin HfO$_2$ insulator  %$\approx20 \)$nm
is grown at about a temperature of $35^{\circ}$C via atomic layer deposition, followed by the deposition of % the metal
$5\,$nm Ti /$150\,$nm Au. Using the same insulator on reference Hall bar structures it is possible to tune the density from $1 \times 10^{12} \, \textrm{cm}^{-2}$ n-type regime to $- 1 \times 10^{12} \, \textrm{cm}^{-2}$ p-type dominated conductance. A false color SEM picture of a final device without an applied gate is shown in Fig. \ref{Fig:fig1}c).

For the transport studies the samples are then cooled down in a dilution refrigerator with a base temperature of $30 \,$mK (Device 1) or $120\,$mK (Device 2-4) and the differential conductance $dI/dV$ is measured using low excitation and low-frequency lock-in techniques combined with DC measurements as depicted in Fig. \ref{Fig:fig1}b). Several devices made from different wafers with and without a top-gate have been measured, yielding all very similar results from which four exemplary devices are discussed. %in the following section.

\section{Method of analysis}
{In the design of the experiment, we anticipate that the transport from N to S will be controlled by the process of Andreev reflection, which allows using the theory of  Blonder $et~al.$ \cite{Blonder1982b} (BTK-theory). This theory assumes thermal equilibrium for the relevant states. %In a S-c-S system the focus is on the $V=0$ Josephson effect. In a N-c-S system the focus is on the voltage dependence of the current, which provides information on the energy-dependence of the states in $S_{\textrm{p}}$ the (proximitized) superconductor. 
In the experimental configuration used by us the occupation of states will potentially deviate from the equilibrium Fermi-Dirac distribution. As shown in Fig. \ref{Fig:fig1}a) we define three sections through which the transport occurs in our device. In that drawing, the wide uncovered part of the 3DTI constitutes the N side and fulfills the criterion of a proper Landauer-B\"uttiker equilibrium reservoir with a Fermi-function at the bath temperature $T_b$ and a Fermi-level $\mu_\textrm{N}=\mu_{\textrm{p1}}-eV_{\textrm{SN}}$, which depends on the applied applied bias $V_{\textrm{SN}}$. 
On the other side of the constriction, located at $Z_{\textrm{p}}$, the main superconductor S$_\textrm{m}$, niobium, induces superconducting correlations in the 3DTI bar S$_\textrm{p}$. Both superconductors form the same macroscopic quantum state. The current through the sample, assumed to enter from the N-part is carried away as a supercurrent. Therefore, we do not expect a voltage drop beyond $Z_{\textrm{p}}$ and the superconducting side is initially, for zero applied applied bias $V_{\textrm{SN}}=0$, characterized by an equilibrium Fermi-function $\mu_{\textrm{p1}}=\mu_{\textrm{p2}}$ at the bath temperature $T_b$. 
With this starting point we anticipate that the conductance as a function of voltage $V_{\textrm{SN}}$ will, in principle, described by:    
 \begin{equation}
 \label{BTK}
 \begin{aligned}
 I_{SN}&=&\frac{1}{eR_N}  \int_{-\infty}^{+\infty} (f_0(E-eV_{\textrm{SN}},T)-f_0(E,T))\\
 &&[1+A(E,Z)-B(E,Z)] \mathrm{dE},
 \end{aligned}
 \end{equation}
 where $f_0(E,T)$ is the Fermi-Dirac distribution at energy $E$ and temperature $T$. $A(E)$, and $B(E)$ are the probability amplitudes for Andreev and normal reflection of an incident electron from and to the normal reservoir. The normal state resistance $R_N$ is assumed to be the resistance arising from the finite number of modes carried by the cross-section. The voltage drop is located at the orifice with elastic scattering parametrized by $Z=Z_{\textrm{p}}$ as indicated in Fig. \ref{Fig:fig1}a).
 
%  which consi with a chemical potential the Fermi-energy in S$_\textrm{m}$, niobium, indicated by $\mu_{\textrm{p1}}$ is equal to the Fermi-energy in the proximitized 3DTI, $\mu_{\textrm{p1}}$, \emph{i.e.} they form the same macroscopic quantum state and the current through the sample, assumed to enter from the N-part is carried away as a supercurrent. The superconducting side S$_\textrm{p}$ is initially, for $V_{\textrm{SN}}=0$, characterized by an equilibrium Fermi-function at the bath temperature $T_b$. %(a band schematic of this situation is shown in Fig. \ref{Fig:two_barriers}a). 
%In that drawing, the wide uncovered part of the 3DTI %will avoid backscattering of reflected electrons and holes again into the orifice 'a'. Hence, it 
%Therefore, we envision that, starting from $V_{\textrm{SN}}=0$, the BTK-analysis can be applied with N to the left and the superconductor S$_\textrm{p}$ to the right of $Z_{\textrm{p}}$ with the voltage drop located at  $Z_{\textrm{p}}$.  % and  %In addition, the condensate in the 3DTI and the niobium share the same macroscopic phase. The gradient of this phase will drive the supercurrent, taking into account the geometry and Cooper-pair densities. 
%The normal part N of the 3DTI is assumed to constitute also an equilibrium reservoir. These different electrochemical potentials will lead to an electrostatic potential due to the geometry around $Z_{\textrm{p}}$, as is usual for a Sharvin point contact. 
 %As a starting point we make the important assumption that the Fermi-functions are undisturbed by the current flow, but 
We do not know the coefficients $A(E)$ and $B(E)$ \emph{a priori}. They contain the spectral information we are interested in %Following the %In setting up our understanding of the experiment in this way we follow the 
%approach taken by Scheer $et~al.$ \cite{Scheer2000} and Averin $et~al.$ \cite{Averin1995} using atomic scale gold (Au) point contacts with a diffusive proximity-induced superconducting state, the coefficients %. In analyzing their results, Scheer $et~al.$ use the method of Averin and Bardas \cite{Averin1995} to convert the normal ($G$) and anomalous ($F$) Green's functions of the Usadel-theory into the Andreev reflection coefficient $A(E)$ and $B(E)$. In analogy, 
%we assume that in Eq. (\ref{BTK}), $A(E)$ and $B(E)$ 
and are the result of the interaction of the superconductor with the confined bar of the 3DTI with its limited geometry, finite elastic mean free path and finite interfacial transparency $Z_{\textrm{m}}$  \cite{Scheer2000,Averin1995}. In addition it needs to be considered, that the normal part is a 3DTI, where helical surface states dominate the transport \cite{Burset2015,Schaffer2012, Snelder15b}.  %But since we assume that for $V_{\textrm{SN}}=0$ pair-correlations are present, the chemical potentials for the condensate fulfill the condition $\mu_{\textrm{p1}}$\, =\, $\mu_{\textrm{p1}}$ and the induced superconducting state is an equilibrium reservoir. Therefore, we envision that, starting from $V_{\textrm{SN}}=0$,  the BTK-analysis can be applied with N to the left of $Z_{\textrm{p}}$ and the superconductor S$_\textrm{p}$ to the right of $Z_{\textrm{p}}$ with the voltage drop located at  $Z_{\textrm{p}}$. The pair chemical potential $\mu_{\textrm{p1}}$ determines the particle-hole symmetry of the Andreev-reflection process, assumed in Eq. (\ref{BTK}). 
%The main concern of the previous paragraph is how to apply the Landauer-B\"uttiker picture, as implicit in Eq. (\ref{BTK}), with two equilibrium reservoirs and a scattering matrix connecting them to a system with superconducting correlations in the presence of a finite voltage-difference between the two reservoirs. Conventionally, it is assumed that it is clear where the scattering region is separated from the equilibrium reservoirs. This is obvious in the original BTK-configuration of a Sharvin point contact separating 2 bulk reservoirs, which naturally leads to Eq. (\ref{BTK}), with $A(E)$ and $B(E)$ determined from the Bogoliubov-De Gennes equations, ignoring self-consistency. In applying this approach to finite length or geometrically more complex hybrid systems it is usually assumed that the pair-potential $\Delta$ in the scattering region is zero. 

In the covered TI bar we allow for a finite paring-potential $\Delta$, which implies, that the self-consistency equation of the Bogoliubov-De Gennes equations
\begin{equation}
\label{SelfConsistency}
\Delta(\vec r)=V_N\sum_{E>0} v^*(\vec r)u(\vec r)[1-2f_0(E)].
\end{equation}
needs to be fulfilled. The value $\Delta$ depends on the distribution function, which for a driven system may differ from the one assumed for equilibrium reservoirs.  

Hence, we will analyze our data under the assumption that Andreev reflection, due to a finite value of $\Delta$, takes place at $Z_{\textrm{p}}$, which allows us to apply Eq. (\ref{BTK}) to our system with initially, for low voltages, the equilibrium reservoirs taken to be in the normal side and in the proximitized HgTe on the superconducting side with a finite value of $\Delta=\Delta_{\textrm{p}}$, although it does not necessarily  resemble a BCS like density of states. 
}

\begin{figure}[hbtp]
\includegraphics[scale=0.45]{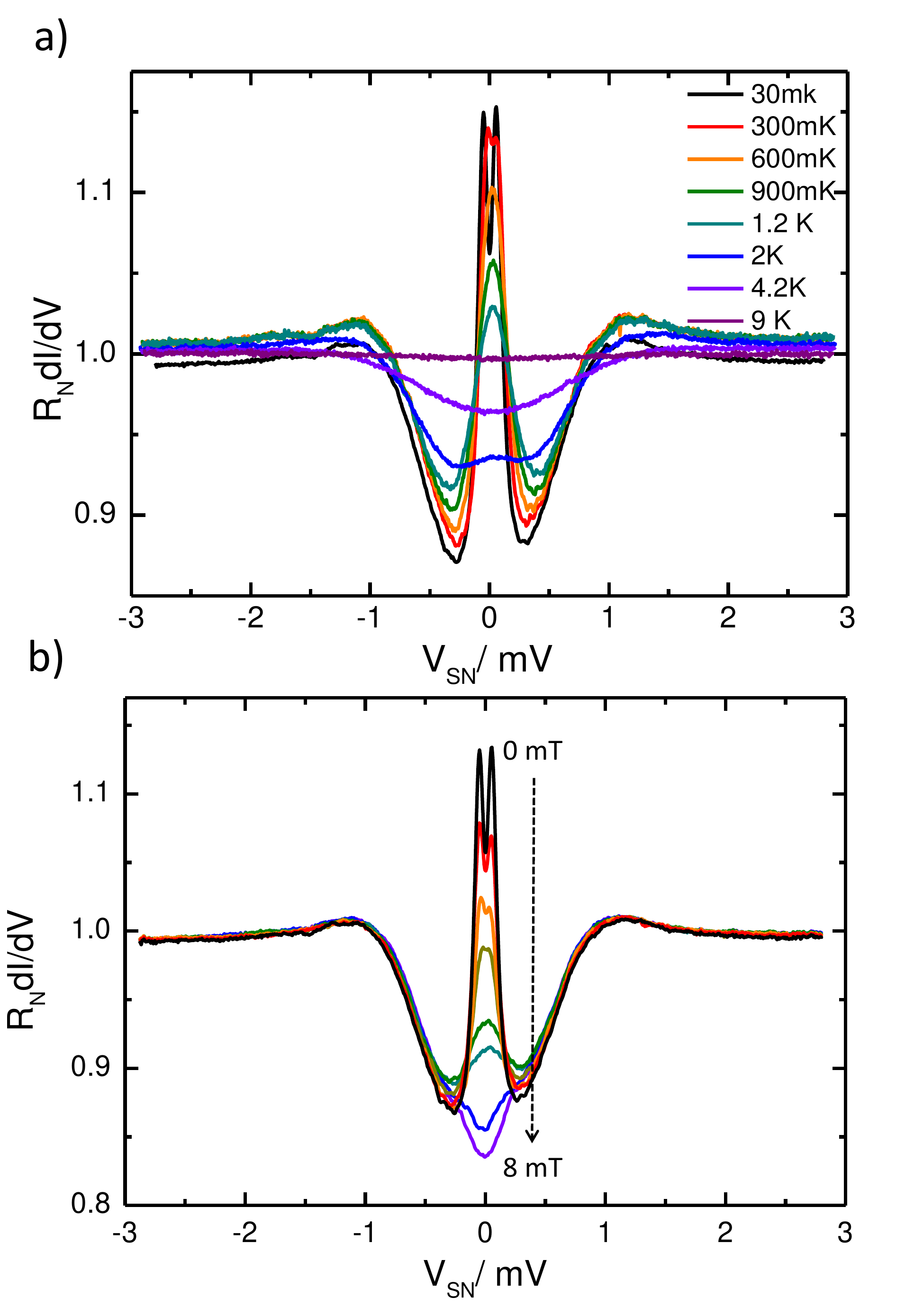}
\caption{a) Conductance of Device 1 normalized to the resistance $R_N$ at $T=9\,$K (purple). At $4.2\,$K an energy gap has clearly opened up due to the niobium being superconducting. Upon lowering the temperature a peak emerges around $V_{\textrm{SN}}=0$, which splits below $500\,$mK. Panel b) shows the conductance measured at $30\,$mK for increasing (small) magnetic field values. This response is independent of the direction of the applied magnetic field. For clarity a small vertical shift has been removed in the presentation of the data to highlight that the high voltage part of the conductance is immune to these magnetic field strengths.}
\label{Fig:tempdependence}
\end{figure}

\section{\label{sec:level3}Experimental data and interpretation}
Fig. \ref{Fig:fig1}d) gives an overview of the differential conductance across the point contact for four different devices at zero applied gate voltage and zero magnetic field. At voltages $\lvert V_{\textrm{SN}}\rvert>1.5\,$mV, larger than $\Delta_{\textrm{Nb}}$, the differential conductance is almost constant and a normal state resistance of $R_{\textrm{N}} = 160-240 \, \Omega$ is observed, depending on the measured device. For voltages around $V_{\textrm{SN}} \simeq 1.1\,$meV the conductance is slightly enhanced which is indicated by the black arrows and then starts to decrease for smaller voltages. Close to zero bias, the conductance enhances again resulting in a double peak structure around $V_{\textrm{SN}}=0$, with a peak separation of about $100\,$\textmu V for Device 1, and slightly different for the other devices. The red arrows are used to draw the attentions to a sample dependent sub gap feature. The four devices differ with respect to the shape and length of the HgTe bar underneath the superconductor. Device 1 is symmetric with width $w=1\,$\textmu m  and two open ends. Device 2 has a step like shape with partially width $w$ and partially width of $0.6\,$\textmu m. Similarly, Device 3 but with the wide 'normal' electrode connected to the wide part rather than the more narrow part. Finally, Device 4 is terminated half-way and implies a largely closed HgTe bar.  At present it is not clear whether this should be interpreted as a feature in the relevant non-equilibrium distribution entering Eq. (\ref{BTK}) or as reflecting a finite size effect of the HgTe in the spirit of the analysis of Kopnin and Melnikov \cite{Kopnin2011a}. Systematic shape-dependent experiments are needed to map and evaluate this dependence accurately and to test the full hypothesis. 
 An asymmetric background for negative and positive bias is observed in all devices. The data can be normalized by multiplying with the normal state resistance $R_N$ measured at $T>T_c$, as shown in Fig. \ref{Fig:tempdependence}, to eliminate this slope. 
We will discuss the observed behavior now in more detail.
 \begin{figure*}
\includegraphics[scale=0.6]{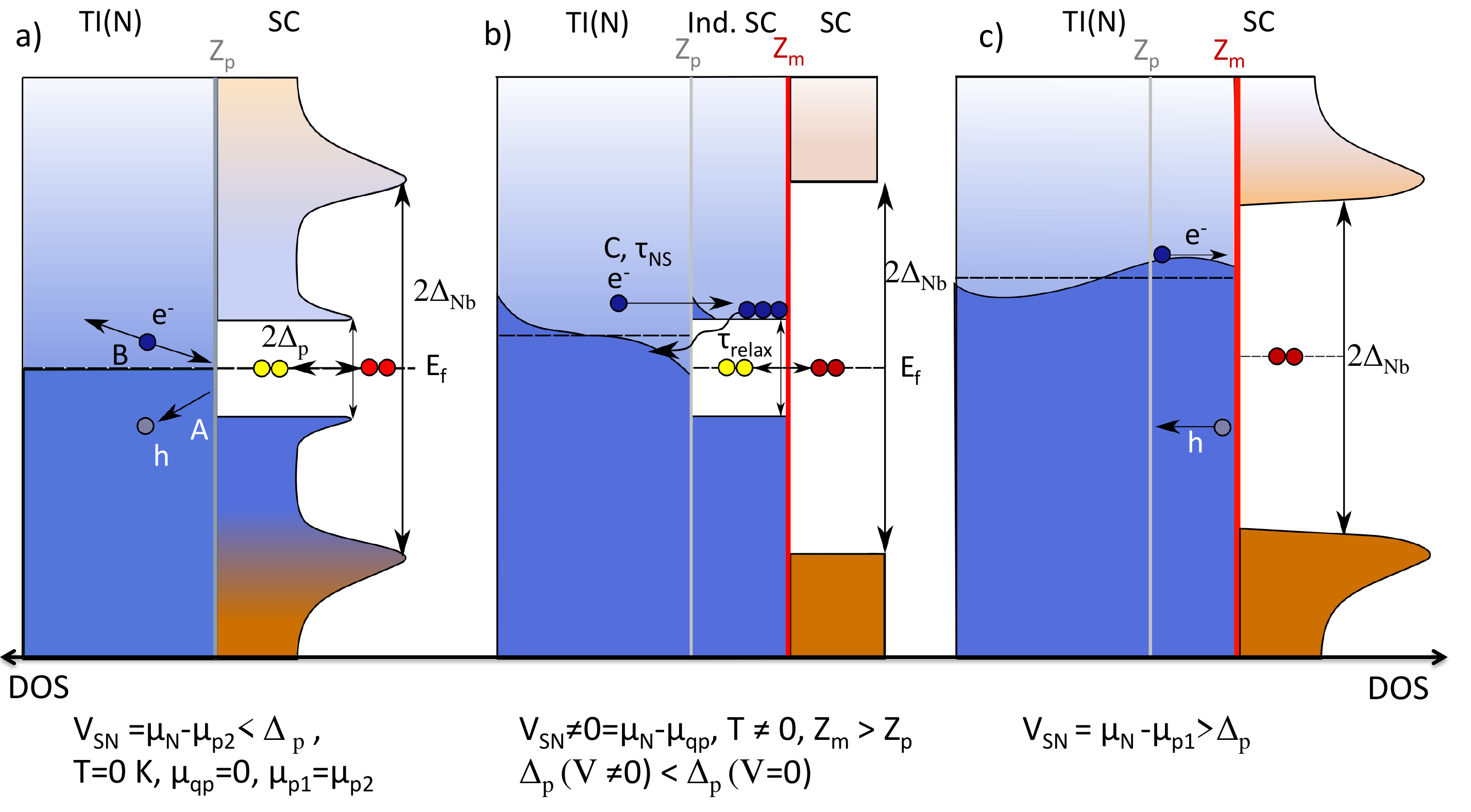}
\caption{Semiconductor representation of the NcS system for different bias regimes. The normal part is the topological insulator, the constriction is, initially characterized by a barrier $Z_{\textrm{p}}$ and the superconductor by a pair potential $\Delta_{\textrm{p}}$. In a) the system is at zero bias and zero temperature. The voltage difference will emerge at the narrow point contact and Andreev reflections (normal reflections) are occurring there with probability A (B). The  TI-Cooper pairs are phase-coherently coupled to the Nb-condensate ($\Delta_{\textrm{Nb}}$). They form one superconducting condensate. In b) at finite temperature and bias electrons from higher energies are allowed to enter the proximity-induced superconductor. These 'hot' electrons are trapped in the proximitized area, because Andreev-reflection does not carry entropy.  The only relaxation mechanism is by electron-phonon relaxation or by contact with a thermal equilibrium reservoir. So the proximity-induced superconducting state S$_\textrm{p}$ is quenched ($\Delta_{\textrm{p}}\rightarrow 0$) and the situation as depicted in c) is present. Then, at higher voltages transport is measured between a normal reservoir being the 3DTI HgTe and the superconductor niobium with an interface resistance characterized by $Z_{\textrm{m}}$.}
\label{Fig:two_barriers}
\end{figure*}
\subsection{Low voltage data: proximity-induced order parameter}
Close to zero bias, we find a strongly enhanced conductance with a double-peak structure in Devices 1-3 and a single peak for the closed bar (Device 4). As shown in Fig. \ref{Fig:tempdependence}, this double-peak structure merges at higher temperatures to a bell-shaped curve. 

The data in this figure is taken for Device 1, which we will focus on for a detailed analysis.  The conductance is  normalized with the normal state resistance $R_N$ above the critical temperature $T>T_c$. % which reflects device-conditions which we seek to elucidate. 
From Fig. \ref{Fig:tempdependence}a) it is clear that at $4.2\,$K an energy gap opens up, which is on the scale of the superconducting niobium gap. Upon lowering the temperature a peak emerges around $V_{\textrm{SN}}=0$, which splits in two below $500\,$mK. 
 
Panel b) of Fig. \ref{Fig:tempdependence} shows the conductance measured at $30\,$mK for increasing values of magnetic field applied perpendicular to the sample. We verified that the response is independent of the direction of the applied magnetic field. For clarity, a small vertical shift has been removed in the presentation of the data in Fig. \ref{Fig:tempdependence}b) to highlight that the high voltage part of the conductance is immune to these magnetic field strengths. Evidently, the central peak can be suppressed completely by applying a magnetic field.  We attribute this central bell-shaped peak, which evolves into a two peak structure as a manifestation of the proximity-induced superconducting order parameter as given by Eq. (\ref{SelfConsistency}).

For $V_{\textrm{SN}}=0$ the system is in equilibrium and the apparent $\Delta$ is the result of electrons in the HgTe bar underneath the niobium film, which are confined in a certain width and length. Their occupation is given by a Fermi-Dirac distribution and it is part of the superconducting equilibrium reservoir S$_\textrm{p}$. For finite voltage bias, the current in S$_\textrm{p}$ is carried away as a supercurrent, and, importantly, the voltage-drop occurring at the interface indicated by $Z{\textrm{p}}$ in Fig. \ref{Fig:two_barriers}a), is due to the difference in electrochemical potentials between N on the left of $Z_{\textrm{p}}$ and S$_{\textrm{p}}$ on the right of $Z_{\textrm{p}}$. The scale of the relevant Sharvin resistance is controlled by the number of modes at the $Z_{\textrm{p}}$ location and by the unknown value of $Z_{\textrm{p}}$. Therefore, the normalization on $R_N$ as defined above is not viable in this equilibrium regime. 

\begin{figure}[hbtp]
\includegraphics[scale=0.45]{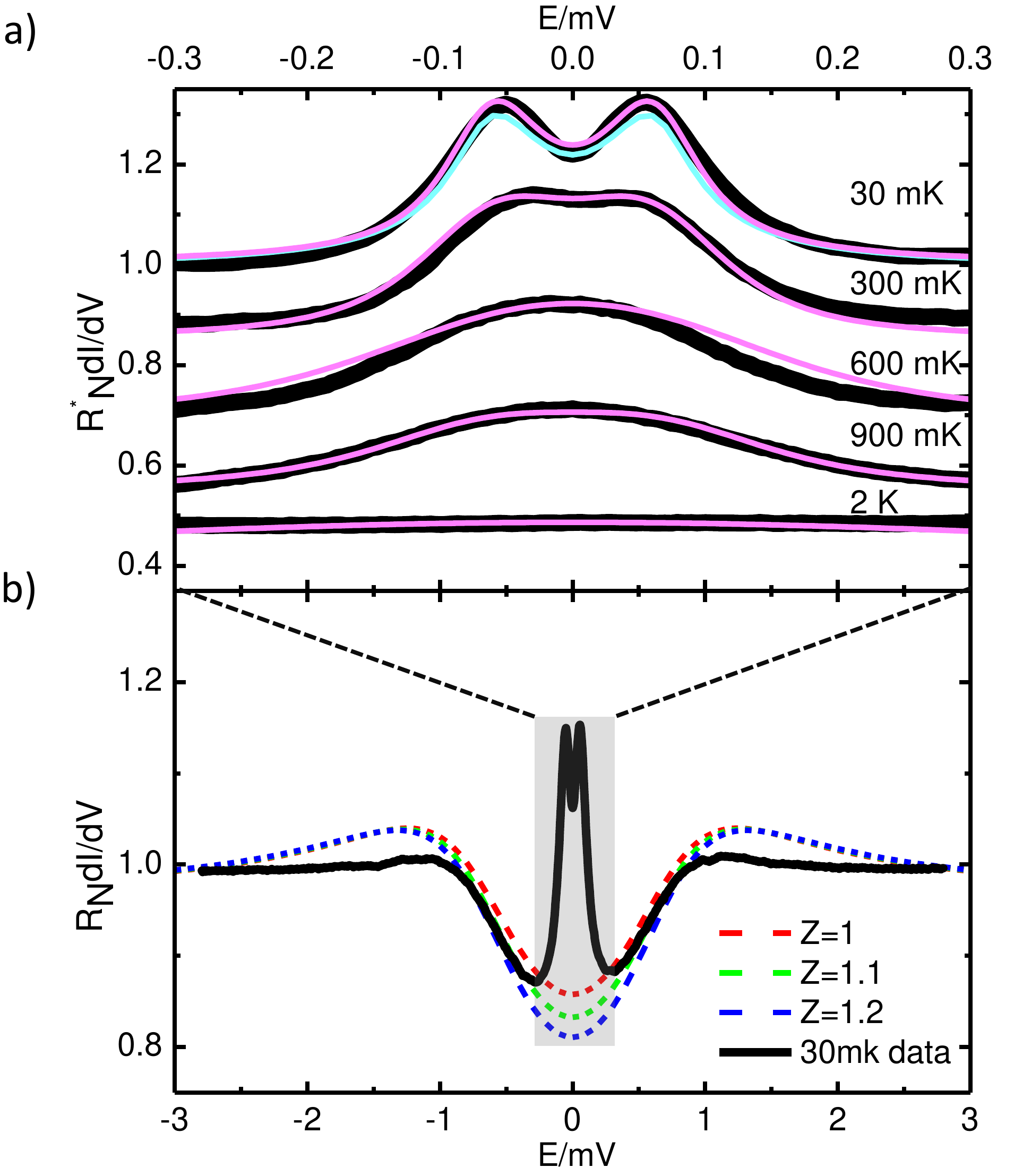}
\caption{In a) the central split peak (gray zone of b)) is compared to an analysis using Eq. (\ref{BTK}) (cyan) with a fixed value of $Z_{\textrm{p}}=0.4$ and a broadening parameter $\Gamma \approx 0.025\Delta_{\textrm{p}}$. The magenta lines show a comparison with the model developed in Ref. \cite{Burset2015} with a broadening parameter $\Gamma<0.015\Delta_{\textrm{p}}$. The value of $\Delta_{\textrm{p}}$ in both models is $70\,$\textmu eV. In panel a) we have abandoned the normalization of the data on $R_N$ at high voltages and in the normal state. Instead we have chosen to take the conductance value at the edge of the gray zone. The precise value is a bit arbitrary, but should be close to this value. The curves are offset for better visibility. b) Conductance of Device 1 normalized with the normal state resistance $R_N$ above the critical temperature $T>T_c$ at $30\,$mK. The gray area indicates the voltage-range where we assume an equilibrium proximity-induced superconducting state. The dashed lines show fits using Eq. (\ref{BTK}) for three different $Z_{\textrm{m}}$ parameters and a broadening of $0.7\Delta_{\textrm{Nb}}$. }
\label{Fig:BTKZ1}
\end{figure}

{In Fig. \ref{Fig:BTKZ1}a), data for different temperatures are compared with standard BTK-modeling using Eq. (\ref{BTK}) (cyan) and the model that explicitly takes the surface states of a 3DTI into account from Ref. \cite{Burset2015} (magenta) both leading to very reasonable agreements. In this figure we have renormalized the data differently.  We have chosen the conductance value at the edge of the gray zone in Fig. \ref{Fig:BTKZ1}b), as a reasonable approximation to the real value of $R_N$ entering Eq. (\ref{BTK}). From the comparison shown in Fig. \ref{Fig:BTKZ1}a), we conclude that we find a proximity-induced order parameter $\Delta_{\textrm{p}}= 70\,$\textmu eV for both models. 

The fits using Eq. (\ref{BTK}) were obtained with a small barrier height $Z_{\textrm{p}}=0.4$. Here, we have assumed that the proximity-induced order parameter $\Delta_{\textrm{p}}$ leads to a standard BCS like behavior of the coefficients $A(E)$ and $B(E)$ as a function of energy and that the normal state is described by a parabolic band dispersion. The model might therefore not capture the microscopic details but makes it suitable to compare to other systems. 

The treatment of Ref. \cite{Burset2015} models the conductance of a NS junction on the surface of a 3DTI, exactly as appropriate for our experiment. The contact between the normal region and the induced superconducting reservoir is modeled as a square potential barrier, where the dimensionless barrier strength $Z_{\textrm{p}}$ is defined as the product of the barrier height and width. %(see Appendix A). 
The sub-gap tunnel conductance of the NS junction is then an oscillatory function of the barrier strength $Z_{\textrm{p}}$ and minimum for values $Z_{\textrm{p}}=(n+1/2)\pi$, with $n$ an integer\cite{Sengupta_2006, Linder_2008}. By applying this model to our experimental data, a rather large barrier can be used. The enhanced conductance can then be seen as a signature of the helical surface states where highly transparent modes are always expected due to Klein tunneling. 

We interpret the low voltage data as a probe of the induced superconducting state in the 3DTI of strained HgTe. There is no reason to expect \emph{a priori} a s-wave order parameter. In fact we expect deviations, such as for example predicted by Burset $et~al.$ \cite{Burset2015}. Since the actual spectra depend on several parameters, a larger data-set is needed to provide a reliable analysis to show the influence of the helical Dirac nature of the surface states. %Especially the values of the barriers and broadening parameters can deviate from the real values, which have to be combined with the expected $A$ and $B$ parameters.
Nevertheless, this open question does not affect the conclusion that we can draw with respect to the identification of the regime, where spectroscopy of the induced superconducting state can reliably be performed. }

\subsection{High voltage data: niobium order parameter}

For voltages larger than $0.5\,$meV, the conductance curves in Fig. \ref{Fig:tempdependence}b) all superimpose, if we excerpt the central part interpreted as the proximity-induced order parameter. The data outside the central part can no longer be interpreted as the conductance of a NcS point contact at $Z_{\textrm{p}}$. The electronic states in the HgTe bar underneath the niobium are no longer correlated as expressed in Eq. (\ref{SelfConsistency}).  
For increasing voltage at the location $Z_{\textrm{p}}$, higher energy quasiparticles are injected into the HgTe bar as depicted in Fig. \ref{Fig:two_barriers}b). They cannot escape into an equilibrium reservoir because of the large gap of the superconductor niobium and {the fact that Andreev reflections do not exchange heat}. Therefore, $f_0(E)$ in Eq. (\ref{SelfConsistency}) 
becomes a non-equilibrium distribution with relatively hot electrons, which leads in general to a destruction of the proximity-induced order parameter $\Delta_{\textrm{p}}$, in the same way as a small magnetic field quenches this induced superconducting state. Hence, beyond a voltage of about $0.5\,$meV the system has changed and we are left with a non-superconducting HgTe bar in contact with niobium (as shown in Fig. \ref{Fig:two_barriers}c) with an interface with an unknown transmissivity parametrized by $Z_{\textrm{m}}$.

The change in conductance around $1.1\,$mV is now naturally attributed to the superconducting gap of the niobium film. The conductance increases slightly, as expected at the superconducting gap edge. For smaller voltages the conductance reduces, an indication of dominant normal reflections over Andreev reflections ($B/A>1$). As shown in Fig. \ref{Fig:BTKZ1}b), we are able to achieve fairly good qualitative agreement with a BTK-analysis as well for this outer gap, using a quite large barrier $Z_{\textrm{m}}  = 1.1$ and $\Delta_{\textrm{Nb}} = 0.8\,$meV, indicating a relatively low transparency of the Nb/HgTe interface. We also need to use a relatively large broadening parameter $\Gamma=0.7\Delta_{\textrm{Nb}}$ which could be caused by the large contact area and spatial gradients at the Nb/HgTe interface. 

%In the previous two sections \ref{sec:level3} A and \ref{sec:level3} B we have used the standard scattering amplitudes of BTK-theory. Based on our analysis of the different voltage regimes this justified in Section \ref{sec:level3} B, because it appears as a probe of the conventional superconductor niobium. In Section \ref{sec:level3} A we interpret the data as a probe of the induced superconducting state in the 3DTI of strained HgTe. There is no reason to expect a priori a s-wave order parameter. In fact we expect deviations, such as for example calculated by Burset $et~al.$ \cite{Burset2014}. Since the actual spectra depend on several parameters, a larger data-set is needed to provide a reliable analysis to show the influence of the helical Dirac nature of the surface states. Especially the values of the barriers and broadening parameters can deviate from the real values, which have to be combined with the expected $A$ and $B$ parameters. Nevertheless, this open question does not affect the conclusion that we can draw with respect to the identification of the regime, where spectroscopy of the induced superconducting state can reliably be performed.

\section{\label{sec:level5}Gate dependence of the conductance\protect}

\textbf{
\begin{figure}
\includegraphics[scale=0.33]{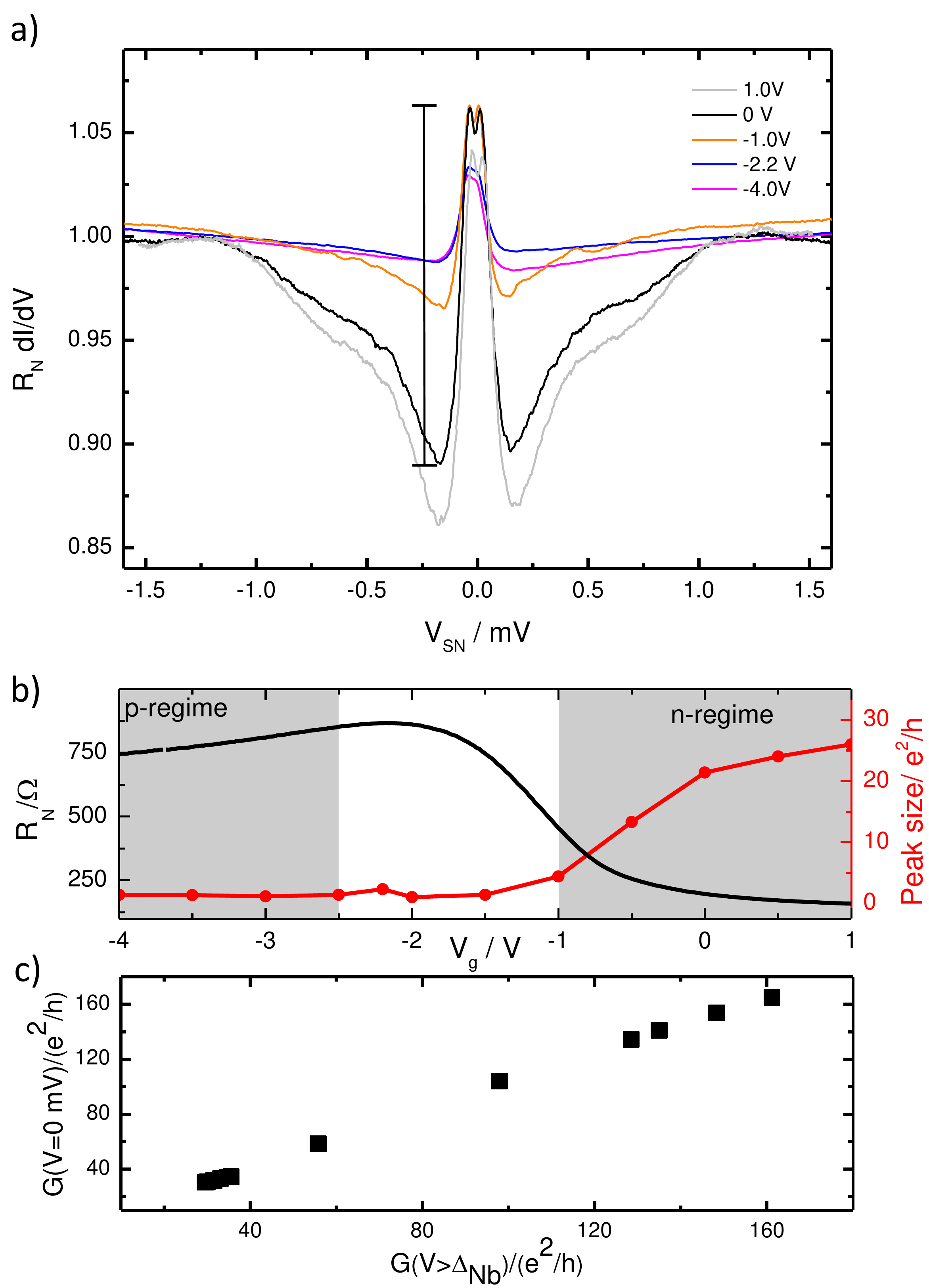}
\caption{a) Gate dependence of normalized conductance of Device 2 at $B=0\,$T from $1\,$V to $-4\,$V. The black bar indicates how the height of the central peak is evaluated in panel b). b) Normal state Resistance $R_N$ versus gate voltage (black) and size of the peak (red) defined as indicated in a) by the black bar for $V_g=0$. c) Normal state conductance versus zero bias conductance is shown.}
\label{Fig:gate}
\end{figure}}

The previous data are all obtained on the electron side (n-type) in which the mobility is high. In Fig. \ref{Fig:gate}a) conductance data are shown for different gate voltages from $+1~V$ to $-4~V$ in which the 3DTI changes from n- to p-type conduction. 
The curves are normalized to the resistance $R_N(T>T_c)$ for each gate voltage individually. %The normal state resistance for  value is shown in Fig.\ref{Fig:gate}b. It follows the standard dependence of the 2-point resistance of strained HgTe in the TI-regime. 
The behavior of the normal state resistance of Device 2 versus the gate voltage is comparable to the reference Hall-bar where we are able to tune the density from initially n-doped, over the charge neutrality point into the hole dominated regime. We distinguish two regimes from $1\,$V to about $-1\,$V the device is in the n-conducting regime. In this regime the mobility is high and the point contact is expected to be ballistic. By tuning into the p-regime the mobility reduces by about a factor of ten  and the mean free path is now smaller than the size of the point contact and, therefore, expected to be in the diffusive regime. 

From the conductance curves (Fig. \ref{Fig:gate}a) it is clear that we observe no longer a signature of the niobium pairing potential in the p-regime. Upon changing the gate voltage, features at the scale of the niobium gap disappear upon approaching the Dirac point (at $-2.2~V$). The only significant voltage-dependent feature is around $\pm~ 100\,$\textmu eV. We assume that this observation is a signature that the NcS point contact is probing the induced superconducting state of the HgTe bar in a diffusive proximity-system, leading to a mini-gap. The height of the zero bias anomaly as a function of gate voltage is quantified using Fig. \ref{Fig:gate}a), by defining $dI/dV_{T=30\,\textrm{mK}}-dI/dV_{T>T_c}$, and plotted in Fig.  \ref{Fig:gate}b) as red dots. The amplitude is several tens of $e^2/h$ in the n-conducting regime and decreases continuously up to the the maximum in resistance region where it saturates at a value of 1-2 $e^2/h$ depending on the sample. %\textit{These considerations about the definition of the peak-height are important, because they could falsely connect to the possible regime of perfect Andreev reflection caused by a Majorana zero mode in the HgTe bar. Another, often used, approach is to scale the conductance at zero bias with the normal state conductance above $\Delta_{\textrm{Nb}}$, for different gate voltages, as shown in Fig. \ref{Fig:gate}c. An almost linear behavior is found with a slope around one indicating that height of the peak scales with the normal state resistance. However, we believe this scaling should be based on the inferred resistance in the normal state at zero bias relevant for the end-point of the HgTe bar.}  %Another interesting observation is that the asymmetry between the positive and negative biases which is present in all devices changes sign which is visible if one compares the curves for $1\,$V and $-4\,$V. This asymmetry is only visible when superconductivity is present. 

\section{General remark about our analysis} 
The analysis of our data has lead us to discuss the conductance data resulting from the transport through three different electron systems (N, S$_\textrm{p}$ and S$_\textrm{m}$), separated by two interfaces of transparency $Z_{\textrm{p}}$ and $Z_{\textrm{m}}$. 
Following Beenakker \cite{Beenakker1992} it is assumed that any contact between a normal reservoir and a superconducting reservoir is given by
\begin{equation}
\label{BeenakkerBTK}
G_S=2G_0\frac{G_N^2}{(2G_0-G_N)^2}
\end{equation}
with $G_0=2e^2/h$ the quantum unit of conductance, $G_N$ the conductance in the normal state, and $G_S$ the conductance with one of the electrodes superconducting. This expression is the zero-voltage limit of the classical BTK-formula for different values of transmissivity $Z$. In order to calculate $G_S$, often the conductance at $V>\Delta_s$ is used as $G_N$ (see also Fig.\ref{Fig:gate}c)) and implying that this experimental value is independent of the applied bias. The most important implication in our case is that one measures at high voltages not a proximity-induced superconducting gap, but rather the parent superconductor. We suggest that the low voltage data should be understood by acknowledging that the scattering region and the equilibrium reservoirs at $V_{\textrm{SN}}=0$ should be defined differently from the one at higher voltages, such as in our case $V>0.8\,$meV. This distinction is in general not specific to our case but should apply to other topological systems, for example the one studied in Kjaergard $et~al.$ \cite{Kjaergaard2017} and Suominen $et~al.$ \cite{Suominen2017} and might explain deviations from expected behavior in these two papers.

\section{Conclusions}
In conclusion, we have carried out transport spectroscopy of the proximity-induced pair-potential of a niobium covered bar of strained HgTe, which has been demonstrated to be prone to be a 3DTI. In analyzing the data we allow for a finite pairing potential in the strained HgTe, in contrast to a commonly made quantum transport simplification as introduced by Lambert \cite{Lambert1991} and Beenakker \cite{Beenakker1992}, in which the properties are assumed to be controlled exclusively by the scattering in the structure. In addition, we take into account how to identify the relevant distribution function over the energies, implying the relevance of a non-equilibrium distribution function in analyzing the data. These results are an important step towards a better understanding and engineering of topological superconductivity and may serve as a building block for a further analysis of the $4\pi$-Josephson effect as reported in Refs. \cite{Wiedenmann2016a,Bocquillon2016b,Deacon2016}.  

\acknowledgments
We like to thank A. Akhmerov, W. Belzig, F.S. Bergeret, and B. Trauzettel for many helpful discussions. The work at W\"{u}rzburg was supported by was financially supported by the German Research Foundation DFG via
SFB 1170 \textquotedblleft ToCoTronics \textquotedblright and the SPP 1666, the Land of Bavaria (Institute
for Topological Insulators and the Elitenetzwerk Bayern and the European Research Council (advanced grant project
3-TOP and 4-TOPS). TMK, who acknowledges support from the European Research Council
Advanced grant no. 339306 (METIQUM), by the Ministry of Education and Science of the Russian Federation, contract 14.B25.31.0007 of 26 June 2013 and TMK and EB thank the Alexander von Humboldt Stiftung. P.B. acknowledges support from the European Union's Marie Sk\l odowska-Curie grant agreement No. 743884. 
\appendix

%\section{Appendixes}

%\begin{figure}[hbtp]
%\includegraphics[width=\linewidth]{SupplementartyFigGate}
%\caption{Gate dependence of reference Hall-bar. Black is the  longitudial resistance in red the mobility. The inset shows the mean free path as a function of gate voltage.}
%\end{figure}
%\begin{figure}[hbtp]
%\includegraphics[width=\linewidth]{appendix2}
%\caption{Longitudinal and transversal resistance of the strained HgTe layers at zero gate voltage and $4\,$K. }
%\label{charac}
%\end{figure}

\appendix

\begin{verbatim}

\end{verbatim}

%\newpage %Just because of unusual number of tables stacked at end
\bibliographystyle{apsrev4-1}
%\bibliography{apssamp_re}


\end{document}